\documentclass{aastex}          
\usepackage{spr-astr-addons}    
\usepackage{graphicx,times}             
\usepackage[caption=false]{subfig}
\usepackage{natbib}
\usepackage{amssymb,amsmath}
\usepackage{float}
\usepackage{url}
\usepackage{upgreek}
\newcommand{\bs}[1]{\boldsymbol{#1}}
\newcommand{\df}{\buildrel \Delta \over = }
\newcommand{\bx}{_{\bs{x}}}

\begin{document}
%
\title{Systematic Low-Thrust Trajectory Optimization for a Multi-Rendezvous Mission using Adjoint Scaling}

\shorttitle{Low-Thrust Trajectory Optimization}
\shortauthors{Tang et al.}

\author{Fanghua Jiang} \and \author{Gao Tang}
\affil{School of Aerospace Engineering, Tsinghua University}
\begin{abstract}
A deep-space exploration mission with low-thrust propulsion to rendezvous with multiple asteroids is investigated. 
Indirect methods, based on the optimal control theory, are implemented to optimize the fuel consumption. 
The application of indirect methods for optimizing low-thrust trajectories between two asteroids is briefly given. 
An effective method is proposed to provide initial guesses for transfers between close near-circular near-coplanar orbits. 
The conditions for optimality of a multi-asteroid rendezvous mission are determined. 
The intuitive method of splitting the trajectories into several legs that are solved sequentially is applied first. 
Then the results are patched together by a scaling method to provide a tentative guess for optimizing the whole trajectory. 
Numerical examples of optimizing three probe exploration sequences that contain a dozen asteroids each demonstrate the validity and efficiency of these methods. 
\end{abstract}


\section{Introduction}
Small bodies in the solar system, especially asteroids, have attracted the attention of both space agencies and scientists for several decades. 
Innovative applications such as deflecting the Earth-crossing asteroids with new propulsion systems such as low-thrust propulsion or solar sail have been widely studied in literature \citep{casalino2012indirect, mckay2011survey, zeng2011new, zeng2014fast, wu2014artificial,gong2015equilibria,mcinnes2002astronomical}. 
Low-thrust propulsion is especially ideal for deep-space missions because of its high specific impulse. 
Its successful application in deep-space missions such as DAWN \citep{DAWN} has demonstrated its capability to increase the payload. 
Fuel-optimal low-thrust trajectories optimization problems are much more difficult to solve because the low thrust leads to a long firing of the engine. 
Interplanetary missions designed to explore multiple targets promote the scientific return and decrease the average expense but also lead to greater challenges in optimizing low-thrust trajectories. 
In this paper, we develop a systematic approach which is used to find the fuel-optimal trajectories of a multi-asteroid rendezvous mission.  

Indirect methods for optimizing low-thrust trajectories can take advantage of the calculus of variation and thus convert the optimal control problem into a multi-point boundary value problem (MPBVP) \citep{Bryson1975}. 
Indirect methods are favored for their efficiency and optimality if proper initial guesses are given. 
The homotopic approach \citep{bertrand2002new, jiang2012}, normalization of initial adjoint variables \citep{jiang2012}, and the switching detection methods \citep{tang2016Capture} are widely applied to overcome the difficulty arising from the bang-bang control. 

In the problem under discussion, transfers preferably take place between close near-circular and near-coplanar orbits, so reasonable simplifications are applied and the closed-form energy optimal transfer is solved analytically to guess the initial adjoint variables. 
Compared with random guesses, providing initial guesses using this method is more reliable and effective. 
The simplifications are based on \citet{casalino2014approximate, gatto2015fast}. 
Similar methods have also been used to provide an initial guess for indirect methods \citep{li2012fuel}. 

The whole trajectories of a multi-asteroid rendezvous mission should be optimized in order not to lose optimality. 
Although the optimization of a low-thrust trajectory from one asteroid to the next within a fixed time is relatively easy, the increment of the asteroid number and setting the rendezvous moments free significantly increase the difficulty. 
The intuitive method is to split the mission into multiple legs each of which drops into the transfer from one asteroid to another, denoted as the single-leg transfer. These legs are then solved sequentially. 
\citet{Yang2015837, JiangGTOC5, CasalinoGTOC5} proposed several methods for optimizing similar missions with multiple targets, but none of the missions is optimized in whole so the optimality loses. 
Our contribution is to develop a systematic method for optimizing the whole trajectory in order not to lose optimality. 
The difficulty arising from the large number of variables is overcome by the adjoint scaling technique which provides a tentative guess that is likely to converge because it satisfies most of the boundary conditions. 

This paper is organized as follows: in Section 2, the indirect methods for optimizing single-leg transfers are introduced. The method for guessing initial adjoint variables is described. 
In Section 3, the necessary conditions for optimality when the whole mission is optimized are derived. 
The adjoint scaling technique is proposed to provide an initial guess. 
In Section 4, numerical examples from the 7th Global Trajectory Optimization Competition (GTOC7) are presented to verify the validity of these methods. 
The low-thrust trajectories of three probes which must rendezvous with more than 10 asteroids each are optimized. 
The results show that our methods can save a considerable amount of fuel.
Finally the conclusion is given in Section 5. 
\section{Fuel-Optimal Single-Leg Transfer}
A single-leg transfer denotes a transfer when the spacecraft rendezvous with one asteroid to its rendezvous with another asteroid. 
In this case, MPBVP degenerates into two-point boundary value problem (TPBVP). 
Instead of the position and velocity of the spacecraft, equinoctial elements (EE), denoted as $\boldsymbol{x}$ which is composed of $(p, e_x, e_y, h_x, h_y, L)$, are used to describe the motion of the spacecraft \citep{walker1985set}. 
The dynamical equations are given by
\begin{equation}
\label{eq:DynEEECI}
\left\{\begin{array}{l}
{\boldsymbol{\dot {{x}}}} = {{\boldsymbol{f}}_0}\left( \boldsymbol{x} \right) + {\boldsymbol{M}}\left( {\boldsymbol{x}} \right)\dfrac{{u{T}{\boldsymbol{\alpha }}}}{m}\\
\dot m =  - \dfrac{{u{T}}}{c}
\end{array}\right.
\end{equation}
where $u$ is the thrust ratio within the interval $[0, 1]$; $T$ is the maximal thrust; $m$ is the mass of the spacecraft; the unit vector $\boldsymbol{\alpha}$ denotes the thrust direction; $c=I_{\rm{sp}}g_0$ where $I_{\rm{sp}}$ is the specific impulse and $g_0$ is the gravitational acceleration at sea-level; the details of $\boldsymbol{f}_0$ and $\bs{M}$ can be found in \citet{GaoKluever-409}. 
Such a choice actually contributes to the robustness and efficiency of our algorithm. 
The performance index is
\begin{equation}
\label{eq:SinLegPer}
J=\dfrac{T}{c}\int_{t_0}^{t_f}u{\rm d}t
\end{equation}
where $t_0$ and $t_f$ denote the initial and final moments, respectively. 
The physical meaning of $J$ is the fuel consumption. It should be noted that minimizing $J$ is equivalent to minimizing $-m(t_f)$. 

The application of indirect methods to solve the single-leg transfer can be found in \citet{casalino20071st, bertrand2002new}. 
We refer to \citet{jiang2012, bertrand2002new} for the details of homotopic approaches. 
\citet{jiang2012} proposed the normalization of initial adjoint variables which is used to help provide initial guesses. 
The switching detection method \citep{tang2016Capture} is effective in solving the bang-bang control as long as the homotopic approach provides a good initial guess. 
The combination of these three techniques yields an efficient method for solving fuel-optimal low-thrust trajectories. 
However, the lack of physical meanings for the adjoint variables still leads to a difficulty in providing initial guesses. 
In most cases, we can only guess them randomly so a multiple start technique has to be applied, which significantly reduces the efficiency. 

\subsection{Guessing Initial Adjoint Variables}
In the problem under investigation most transfers, at least the preferable ones, are between close, near-circular and near-coplanar orbits. 
\citet{casalino2014approximate} investigated time-optimal transfers between close low-eccentricity orbits with little change of inclination, which inspired the method proposed here. 
Some reasonable simplifications are applied based on the fact that $e_x, e_y, 
h_x, h_y$ are small and the change of $p$ is also small. 
By simplifying $e_x, e_y, h_x, h_y$ to be 0 and introducing a constant $p'$ which is chosen to be the average of the initial and target orbit, the dynamical equations are simplified as
\begin{equation}
{\boldsymbol{\dot {{x}}}} = {{\boldsymbol{f}}_0'} + \dfrac{uT}{m}{\boldsymbol{M}'}\bs{\alpha}
\end{equation}
where the vector fields $\boldsymbol{f}_0'$ and $\boldsymbol{M}'$ are defined as
\begin{equation}
\label{eq:SimDyn}
\begin{array}{cc}
{{\boldsymbol{f}}'_0} = \sqrt {\dfrac{{{\mu _0}}}{p'}} \left[ 
{\begin{array}{*{20}{c}}
	0&0&0&0&0&	{1/p'}
	\end{array}} \right]^{\rm{T}}\\
{\boldsymbol{M}'} = \sqrt {\dfrac{p'}{{{\mu _0}}}} 
\left[ {\begin{array}{*{20}{c}}
	0&{2p'}&0\\
	{\sin L}&{2\cos L}&{ 0}\\
	{ - \cos L}&{2\sin L }&{0}\\
	0&0&{\cos L/2}\\
	0&0&{\sin L/2}\\
	0&0&0
	\end{array}} \right]
\end{array}
\end{equation}

After the simplification of the dynamical equations, it is obvious from equation~\eqref{eq:SimDyn} that $\dot{L}$ is constant during the transfer, which eliminates the possibility of simultaneously satisfying both the change of $t$ and $L$. 
However, this is still acceptable if we only want to generate an initial guess. 
Another reason is that transfers with an improper selection of transfer time and phases are mostly eliminated in the preliminary design. 
The change of mass is neglected, otherwise $\lambda_m$ should be considered, and it would be difficult to obtain a closed-form solution. 
This simplification is reasonable because the high efficiency of low-thrust propulsion leads to a small amount of fuel consumption. 
Another simplification is to assume that ${u}$ is boundless and to seek the energy optimal transfer, otherwise the bang-bang control has to be taken into consideration, for which it is difficult to obtain a closed-form solution. 
The Hamiltonian is built as
\begin{equation}
\label{eq:SimH}
H = {{\boldsymbol{\lambda }\bx^{\rm{T}}}}{{\boldsymbol{f'}}_0} + 
\frac{T}{m}{{\boldsymbol{\lambda }\bx^{\rm{T}}}}{\boldsymbol{M'}\bs{u}} + 
\frac{T}{c}{\boldsymbol{u}} \cdot {\boldsymbol{u}}
\end{equation}
where $\boldsymbol{u}=u\boldsymbol{\alpha}$. Because $H$ does not depend on $p$, $e_x$, $e_y$, $h_x$, $h_y$ (note that $p'$ in $\boldsymbol{M'}$ are chosen to be constant), adjoint variables $\lambda_p, \lambda_{e_x}, \lambda_{e_y}, \lambda_{h_x}, \lambda_{h_y}$ are actually adjoint constants.
The adjoint variable $\lambda_L$ does change during the transfer, but it does not affect the optimal control because the 6th row of $\boldsymbol{M'}$ are all 0. 
As a result, the change in $\lambda_L$ is neglected.

The optimal control which minimizes $H$ is
\begin{equation}
{\boldsymbol{u}} =  - 
\dfrac{{c}}{{2m}}{{\boldsymbol{M'}}^{\rm 
T}}{{\boldsymbol{\lambda}\bx }}.
\end{equation}
With the optimal $\boldsymbol{u}$, the dynamical equation is
\begin{equation}
\label{eq:ChangeofEE}
{\boldsymbol{\dot x}} = {{\boldsymbol{f'}}}_0 - \dfrac{Tc}{{2{m^2}}}{\boldsymbol{M'}}{{{\boldsymbol{M'}}}^{\rm 
T}}{\boldsymbol{\lambda }\bx}.
\end{equation}
Denote $\boldsymbol{N}=\boldsymbol{M'}\boldsymbol{M'}^{\rm{T}}$. It is obvious that the $L$ appearing in $\boldsymbol{N}$ depends on time. 
With the simplification 
\begin{equation}
L=L_0+\omega t
\end{equation}
 where $L_0$ is the $L$ when 
the transfer begins and $\omega=\sqrt{\mu_0/p'^3}$, equation~\eqref{eq:ChangeofEE} is analytically 
integrable and the details are given in APPENDIX A. It is obvious that the changes of $p$, $e_x$, $e_y$, $h_x$, $h_y$ are linear with respect to $\lambda_p, \lambda_{e_x}, \lambda_{e_y}, \lambda_{h_x}, \lambda_{h_y}$. 
With a given orbital transfer problem, the changes of $p$, $e_x$, $e_y$, $h_x$, $h_y$ are known and the corresponding adjoint variables are calculated by solving a system of linear equations. 

The advantage of this method is obviously the high efficiency. However, neglecting the changes in $L$ and $m$ influences the accuracy of the obtained adjoint variables. 
To obtain the initial guess, $\lambda_L$ and $\lambda_m$ are 
guessed randomly in interval $[-1, 1]$ and $[0, 1]$, respectively. 
We apply the technique of normalizing the initial adjoint variables to increase the robustness of single-leg solving by introducing $\lambda_0$ \citep{jiang2012} and setting it to unity. 
A scaling is applied to $\boldsymbol{\lambda}_x$, $\lambda_m$ and $\lambda_0$ so they are on the surface of a high-dimensional sphere after the scaling. 
It is widely known that the change of phase during orbital transfer is fuel-consuming and even a small deviation from the proper phases might lead to a significant increase in fuel consumption. In other word, this method works well for the problems without the constraints of phase such as transfers between two orbits, meanwhile it may fail to deal with rendezvous problem. 
Through preliminary designs the transfers which take place between improper phases are somewhat eliminated so this method should always be the first choice. 
If this method fails to provide an initial guess which eventually leads to convergence, the method of randomly guessing with multiple starts should be applied. 

\section{Fuel-Optimal Multi-Asteroid Transfer}
\begin{figure}[htb]
	\centering\includegraphics[width=\columnwidth]{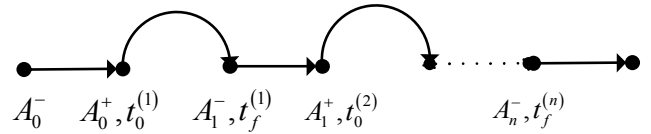}
	\caption{Multi-Asteroid Rendezvous Mission}
	\label{fig:MultiOpt}
\end{figure}

The positive multiplier $\lambda_0$ \citep{jiang2012} is removed from this section because the adjoint variables are not randomly guessed anymore. 
The initial adjoint variables $\bs{\lambda}\bx$ and $\lambda_m$ of every single-leg transfer have to be divided by the corresponding $\lambda_0$ to yield the same optimal control, which is equivalent to a scaling that sets $\lambda_0$ to unity. 
Denote $A_0$, $A_1$, ..., $A_n$ as the sequence of asteroids, $t_0^{(i)}$ the moment 
to leave $A_{i-1}$, $t_f^{(i)}$ the moment to rendezvous with $A_i$, $i=1, ..., n$, 
and ${\Delta}t$ the minimum time to stay on the asteroids. 
It is supposed that $t_0^{(i)}=t_f^{(i-1)}+\Delta t$ for simplicity which indicates that the spacecraft stays for the minimum time at the asteroid. 
The superscripts $+$ and $-$ denote when the spacecraft arrives at and leaves the asteroid, respectively. 
As shown in Figure~\ref{fig:MultiOpt}, a mission containing $n$ single-leg transfers should be optimized. 
In our case, $t_0^{(1)}$ and $t_f^{(n)}$ are fixed while $t_f^{(1)}, t_0^{(2)}, ...,t_0^{(n)}$ are optimized subject to the inequality constraints $t_0^{(i)}<t_f^{(i)}, i=1, ..., n-1$. 
However, these inequality constraints are not imposed but checked a posteriori. 
In fact, $t_0^{(1)}$ is set to match the phase when the spacecraft leaves the Earth, rendezvous with and stays at asteroid $A_0$, and $t_f^{(n)}$ is chosen according to the length of the mission. 

\subsection{Multi-Point Boundary Value Problem}
The EE of the spacecraft at $t_0^{(i)}$ and $t_f^{(i)}$ are constrained to be the same as that of the corresponding asteroid. For an intermediate asteroid $A_i$, $i=1,...,n-1$ the constraints are
\begin{equation}
\label{eq:con_x_tf}
\boldsymbol{x}(t_f^{(i)})-\boldsymbol{x}_{A_i}(t_f^{(i)})=\bs{0}
\end{equation}
\begin{equation}
\label{eq:con_x_t0}
\boldsymbol{x}(t_0^{(i+1)})-\boldsymbol{x}_{A_{i}}(t_0^{(i+1)})=\bs{0}
\end{equation}
\begin{equation}
\label{eq:con_t0_tf}
t_0^{(i+1)}-t_f^{(i)}-\Delta t=0
\end{equation}
\begin{equation}
\label{eq:con_m_t0_tf}
m(t_0^{(i+1)})-m(t_f^{(i)})=0
\end{equation}
These constraints hold only for those $n-1$ intermediate asteroids, i.e. $A_1$ to $A_{n-1}$. 
For $A_0$ and $A_n$ only equation~\eqref{eq:con_x_t0} and \eqref{eq:con_x_tf} hold, respectively. 
To handle these constraints equations~\eqref{eq:con_x_tf}--\eqref{eq:con_m_t0_tf} are multiplied by the numerical adjoint multipliers $\boldsymbol{\chi}_i$, $\boldsymbol{\chi}'_i$, $\chi''_i$, and $\chi_{m_i}$, respectively; the transversality and static conditions at $t_f^{(i)}$ and $t_0^{(i+1)}$ are
\begin{equation}
\label{eq:FreeLmdxtf}
(-\boldsymbol{\lambda}\bx(t_f^{(i)})+\boldsymbol{\chi}_i)=\boldsymbol{0}
\end{equation}
\begin{equation}
\label{eq:FreeLmdxt0}
(\boldsymbol{\lambda}\bx(t_0^{(i+1)})+\boldsymbol{\chi}'_i)=\boldsymbol{0}
\end{equation}
\begin{equation}
\label{eq:H_t_f}
H(t_f^{(i)})-\boldsymbol{\chi}_i\cdot\dot{\boldsymbol{x}}_{A_i}(t_f^{(i)})-\chi''_i=0
\end{equation}
\begin{equation}
\label{eq:H_t_0}
-H(t_0^{(i+1)})-\boldsymbol{\chi}'_i\cdot\dot{\boldsymbol{x}}_{A_i}(t_0^{(i+1)}) +\chi''_i =0
\end{equation}
\begin{equation}
\label{eq:lmdm_t_f}
-\lambda_m(t_f^{(i)})-\chi_{m_i} =0
\end{equation}
\begin{equation}
\label{eq:lmdm_t_0}
\lambda_m(t_0^{(i+1)})+\chi_{m_i} =0
\end{equation}
where only $L$ of ${\boldsymbol{x}}_{A_i}$ depends on time. 
Equations~\eqref{eq:FreeLmdxtf} and \eqref{eq:FreeLmdxt0} suggest the discontinuity of $\boldsymbol{\lambda}\bx$. 
After algebraic manipulations it is easily derived from equations~\eqref{eq:H_t_f} and \eqref{eq:H_t_0} that 
\begin{equation}
\label{eq:Static}
\begin{array}{l}
H(t_f^{(i)})-\lambda_L(t_f^{(i)})\dot{L}_{A_i}(t_f^{(i)})=\\
\qquad H(t_0^{(i+1)})-\lambda_L(t_0^{(i+1)})\dot{L}_{A_i}(t_0^{(i+1)})
\end{array}
\end{equation}
and the combination of equations~\eqref{eq:lmdm_t_f} and \eqref{eq:lmdm_t_0} leads to
\begin{equation}
\label{eq:lmdmcontinue}
\lambda_m(t_0^{(i+1)})=\lambda_m(t_f^{(i)})
\end{equation}
which indicates that $\lambda_m$ is continuous during the mission. As with the single-leg transfer, the mass of the spacecraft at $t_f^{(n)}$ has no constraint so
\begin{equation}
\label{eq:lmdmtf}
\lambda_m(t_f^{(n)})=0.
\end{equation}

In summary, the variables to be solved are: $t_f^{(1)}$ to $t_f^{(n-1)}$, $\boldsymbol{\lambda}_x(t_0^{(1)})$ to $\boldsymbol{\lambda}_x(t_0^{(n)})$, and $\lambda_m(t_0^{(1)})$. The total number of variables to be solved is $7n$. The constraints to be satisfied are: $\boldsymbol{x}(t_f^{(i)})=\boldsymbol{x}_{A_i}(t_f^{(i)}),i=1,...,n$, which gives $6n$ equations; there are $n-1$ static conditions in the form of equation~\eqref{eq:Static} with $i=1,...,n-1$ and $\lambda_m(t_f^{(n)})=0$. 
The MPBVP is built and then solved with shooting methods. 

\subsection{Adjoint Scaling Technique}
The sensitivity of the shooting function, i.e. MPBVP, increases when the number of legs increases. The number of variables to be solved is linear with respect to the number of legs. 
A mission containing many legs is thus difficult to optimize because of the large number of variables to be solved. 
Randomly guessing with multiple starts is not efficient. 
To help guess the initial values of adjoint variables, the mission is split into several legs which are solved sequentially. 
Using the techniques applied to solve the single-leg transfer, every leg is solved efficiently. 
Denote as $\boldsymbol{\Lambda}^{(i)}\df[\boldsymbol{\lambda}\bx(t_0^{(i)});\lambda_m(t_0^{(i)})]$ the solution of leg $i$ when the bang-bang control is solved and $\lambda_0$ is removed. 

The sign of the switching function, denoted as $\rho$ determines whether the thruster is on or off \citep{jiang2012}. It is defined as
\begin{equation}
\label{eq:rho}
\rho  = 1 - {{{\lambda _m}}} - \dfrac{{c}}{{m}}\left\| {{{\boldsymbol{M}}^{\rm{T}}} {{\boldsymbol{\lambda }}\bx}} \right\|.
\end{equation}
where $\boldsymbol{\lambda_x}$ and $\lambda_m$ are the adjoint variables. 
Because multiplying $\rho$ and $\boldsymbol{\lambda}\bx$ by a positive scalar, denoted as $k$, at any instantaneous moment, denoted as $t'$, does not change the sign of $\rho$ (thus the thrust magnitude) or the thrust direction, the optimal control at $t'$ does not change. 
It can be inferred from the dynamical and adjoint differential equations \citep{GaoKluever-409} that $\dot{\rho}$ and $\dot{\bs{\lambda_x}}$ are also multiplied by $k$ while $\dot{\bs{x}}$ and $\dot{m}$ do not change. As a result, the optimal control stays invariant for the whole trajectory. 
It is obvious that the change of $\lambda_m$, denoted as $\Delta\lambda_m$ in a single leg is also multiplied by $k$. 

We might as well investigate two sequential legs whose initial adjoint variables for fuel-optimal transfer are $\boldsymbol{\Lambda}^{(i-1)}$ and $\boldsymbol{\Lambda}^{(i)}$, respectively. 
It is obvious that $\lambda_m^{(i-1)}(t_f^{(i-1)})$ of leg $i-1$ is 0, as is $\lambda_m^{(i)}(t_f^{(i)})$. 
However, $\lambda_m^{(i)}(t_0^{(i)})$ is positive, otherwise the condition $\lambda_m^{(i)}(t_f^{(i)})=0$ cannot be satisfied. 
As a result, equation~\eqref{eq:lmdmcontinue} is not satisfied. 
To fix this error, $\boldsymbol{\lambda}\bx^{(i-1)}(t_0^{(i-1)})$ and $\rho(t_0^{(i-1)})$ are multiplied by $k$, i.e.
\begin{equation}
\label{eq:lmdxatsegmenti_1}
{{\boldsymbol{\lambda }}\bx^{(i-1)'}}(t_0^{(i-1)})=
k{{\boldsymbol{\lambda }}\bx^{(i-1)}}(t_0^{(i-1)})
\end{equation}
\begin{equation}
\label{eq:rhoatsegmenti_1}
\rho^{(i-1)'}(t_0^{(i-1)})=k\rho^{(i-1)}(t_0^{(i-1)})
\end{equation}
to satisfy
\begin{equation}
\label{eq:r}
\lambda_m^{(i-1)'}(t_f^{(i-1)})=\lambda_m^{(i)}(t_0^{(i)})
\end{equation}
where the superscript $'$ means the adjoint variables after the scaling. 
The change of $\lambda_m$ of leg $i-1$ is multiplied by $k$ so $\lambda_m$ at the initial moment of leg $i-1$, denoted as $\lambda_m^{(i-1)'}(t_0^{(i-1)})$, is 
\begin{equation}
\label{eq:lmdmatsegmenti_1}
\lambda_m^{(i-1)'}(t_0^{(i-1)})=k\lambda_m^{(i-1)}(t_0^{(i-1)})+\lambda_m^{(i)}(t_0^{(i)}).
\end{equation}
After algebraic manipulations of equations~\eqref{eq:lmdxatsegmenti_1}--\eqref{eq:lmdmatsegmenti_1} we obtain
\begin{equation}
\label{eq:solofk}
k=1-\lambda_m^{(i)}(t_0^{(i)})
\end{equation}
\emph{Remark}:
The scaling factor $k$ has to be positive which is equivalent to saying that $\lambda_m^{(i)}(t_0^{(i)})$ cannot exceed unity. The adjoint scaling is equivalent to resolving the single leg so the optimal control does not change while $\lambda_m(t_f)$ is set to a new positive scalar, denoted as $\lambda_m'$. 
This is accomplished by choosing the performance index as
\begin{equation}
\label{eq:SinLegPerAdd}
\begin{aligned}
J'&=\dfrac{T}{c}\int_{t_0}^{t_f}u{\rm d}t+\lambda_m'm(t_f)\\
&=m(t_0)+(\lambda_m'-1)m(t_f)
\end{aligned}
\end{equation}
Because the fixed $m(t_0)$ does not influence $J'$, the performance index is equal to $(\lambda_m'-1)m(t_f)$. 
Taking into account that the original performance index is $-m(t_f)$, the two problems yield the same optimal control as long as $\lambda_m'-1<0$, which is why $\lambda_m^{(i)}(t_0^{(i)})$ cannot exceed unity, otherwise the problem becomes maximizing the fuel consumption. 

\begin{figure}[htbp]
	\centering\includegraphics[width=\columnwidth]{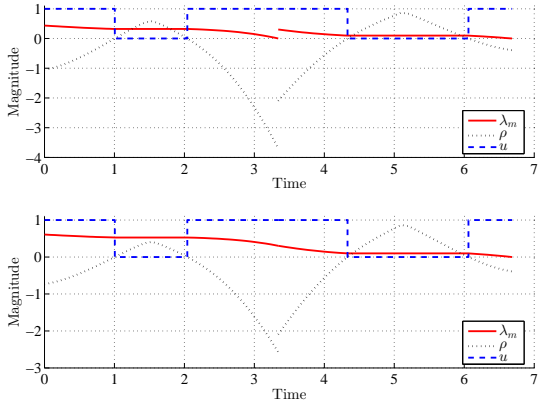}
	\small \caption{History of $\lambda_m$, $\rho$, and $u$ before and after Scaling. Top: before. Bottom: after}
	\label{fig:TransF}
\end{figure}

A brief illustration is shown in Figure~\ref{fig:TransF}. Before the scaling, the results of two fuel-optimal single-leg transfers cannot guarantee the continuity of $\lambda_m$. After the transformation, the control stays invariant but $\lambda_m$ becomes continuous, which is required by the boundary conditions. 

For the mission under investigation which contains more than just two legs, every leg is solved sequentially in the first step, where $t_0^{(1)}$ to $t_f^{(n)}$ are obtained from the preliminary design. The initial adjoint variables of every leg are obtained. 
Then the aforementioned method of scaling the initial adjoint variables is implemented to obtain an initial guess for solving the whole mission. 
The solution of the last leg does not have to be scaled, but the solutions of other legs have to be scaled backwards from leg $n-1$ to leg 1. 
For the $i$-th leg, $\boldsymbol{\lambda}^{(i)}$ is scaled according to $\boldsymbol{\lambda}^{(i+1)'}$ which has been scaled instead of the original one. 
This initial guess satisfies all the boundary conditions except for the static conditions and is more likely to converge than the random guess. 
However, it is possible that at the $i$-th leg, ${\lambda_m^{(i)}}'(t_0^{(i)})$ exceeds unity after the scaling. 
On the condition that $k>0$ is satisfied for every leg, equation~\eqref{eq:lmdmatsegmenti_1} indicates that $\lambda_m^{(i)'}(t_0^{(i)})$ is always increasing when $i$ is decreasing from $n-1$ to 1. 
Although it rarely happens in our numerical examples, it is possible that $\lambda_m'$ exceeds unity and $k$ becomes negative which contradicts the assumption so the optimal control cannot stay invariant. 

However, there is a simple method to estimate whether such problem might happen or not. Equations~\eqref{eq:lmdmatsegmenti_1} and \eqref{eq:solofk} are combined to derive
\begin{equation}
\begin{array}{l}
\lambda_m^{(i-1)'}(t_0^{(i-1)})=(1-\lambda_m^{(i)'}(t_0^{(i)}))\lambda_m^{(i-1)}
(t_0^{(i-1)})\\
+\lambda_m^{(i)'} (t_0^{(i)})=1-(1-\lambda_m^{(i)'}(t_0^{(i)}))(1-\lambda_m^{(i-1)}(t_0^{(i-1)}))
\end{array}
\end{equation}
There are four cases according to the value of $\lambda_m^{(i-1)}(t_0^{(i-1)})$ and $\lambda_m^{(i)'}(t_0^{(i)})$:
\begin{enumerate}
	\item $\lambda_m^{(i-1)}(t_0^{(i-1)})>1$
	\begin{enumerate}
		\item $0<\lambda_m^{(i)'}(t_0^{(i)})<1$: $\lambda_m^{(i-1)'}(t_0^{(i-1)})$ is larger than unity.
		\item $\lambda_m^{(i)'}(t_0^{(i)})>1$: $\lambda_m^{(i-1)'}(t_0^{(i-1)})$ is smaller than unity, but might be negative.
	\end{enumerate}
	\item $0<\lambda_m^{(i-1)}(t_0^{(i-1)})<1$
	\begin{enumerate}
		\item $0<\lambda_m^{(i)'}(t_0^{(i)})<1$: $\lambda_m^{(i-1)'}(t_0^{(i-1)})$ is positive, larger than $\lambda_m^{(i)}(t_0^{(i)})$, and smaller than unity. 
		\item $\lambda_m^{(i)'}(t_0^{(i)})>1$: $\lambda_m^{(i-1)'}(t_0^{(i-1)})$ is larger than unity. 
	\end{enumerate}
\end{enumerate}
As a result, if the original results satisfy $\lambda_m(t_0)<1$ for every leg, $k$ is always positive. 
On the contrary, any leg whose $\lambda_m(t_0)$ exceeds unity will cause the problem. 
There are three methods to handle such a problem: 1) $k$ is chosen to be the same as the former one to avoid possible problems; 2) the preliminary design is refined so such a problem might be avoided; and 3) the adjoint variables are guessed randomly for this leg. 
It is obvious that the first method is the easiest to use but might not lead to convergence. The second method is actually difficult to use and currently there is no method to estimate whether $\lambda_m(t_0)$ will exceed unity. The third method needs to be combined with multiple starts. A method for overcoming such a difficulty is the future work. 

\section{Numerical Examples}
To validate the methods proposed in this paper, three multi-rendezvous sequences in a mission originated from GTOC7\footnote{Available online at \url{http://sophia.estec.esa.int/gtoc_portal/?page_id=515}, retrieved 06 December 2015.} are optimized where every probe rendezvous with dozens of asteroids. 
The topic of GTOC7 is the multi-spacecraft exploration of the main-belt asteroids and the three probes, initially carried by the mother ship, should visit different sequences of asteroids each. 
A brief introduction of the problem is given. 

A mother ship launches from the Earth and releases three probes which must rendezvous with as many asteroids as possible and return to and rendezvous with the mother ship. 
We refer to the website for the details of the mother ship because the trajectory of the mother ship is not considered in this paper. 
Every probe has an electric propulsion system with a specific impulse of 3000 s and a maximum thrust level of 0.3 N. After being released, the probes must return to the mother ship within 6 years. The probe has to stay at every asteroid for at least 30 days. 
Besides the thrust propulsion, the probes suffer only the central gravitation of the Sun. The candidate asteroids move on Keplerian orbits. 
The primary performance index is to maximize the overall number of asteroids reached by the probes. The sum of probe masses when the mission ends is the secondary performance index. 

The first step to solve the problem is to determine the asteroid sequences through the preliminary design. 
The authors' team from Tsinghua University proposed a tree search algorithm with trimming strategy to find the sequences. 
After obtaining the sequences the trajectory of every probe has to be optimized to improve the secondary performance index. 
Every probe starts from a rendezvous with the head of the sequence and finishes the mission when it reaches the tail of the sequence. 
During the competition our team employed the intuitive method of optimizing single-leg transfers sequentially. 
Our final result has a primary performance index of 32 and secondary performance index of 2457 kg. It should be noted that due to the carelessness in programming, one asteroid was visited twice. 
The final ranking\footnote{Data available online at \url{http://sophia.estec.esa.int/gtoc_portal/wp-content/uploads/2014/09/gtoc7_ranks.pdf}, retrieved 06 December 2015.}
of the top five teams is listed in Table~\ref{tab:FinalRank} where $J$ and $J'$ denote the primary and secondary performance index, respectively. 
It is apparent that the secondary performance index is essential to determine the final ranking. 
Our methods are designed for improving the secondary performance index. 
\begin{table}[htbp]
	\caption{Ranking of Top Five Teams of GTOC7}
	\label{tab:FinalRank}
	\centering
	\begin{tabular}{c | c | c | c } 
		\hline
		Rank & Team & $J$ & $J'$ \\
		\hline
		1 &  JPL 		  & 36 & 2450.3 \\
		2 &  ACT/ESA-ISAS & 35 & 2502.2 \\
		3 &  Un. Texas 	  & 35 & 2493.0 \\
		4 &  CAS & 32 & 2509.7 \\
		5 &  Tsinghua Un. & 32 & 2457.0 \\
		\hline
	\end{tabular}
\end{table}

All the computations are executed on a desktop personal computer with a CPU of 3.60 GHz. The programs are written in C++ and compiled with Microsoft Visual Studio Express 2013. 
All quantities concerning the length are nondimensionalized with the astronomical unit (AU, $1.49597870691 \times 10^8 \rm{km}$); the time is so nondimensionalized that the angular velocity of a circular orbit whose radius is 1 AU is unity; the gravitational parameter of the Sun is nondimensionalized to unity; and the mass of the spacecraft is nondimensionalized with its initial mass. 

We need four steps to obtain the fuel-optimal bang-bang control. 
First, multiple single-leg transfers are solved sequentially. 
The initial adjoint variables are guessed with the aforementioned method, and they can all lead to convergence. 
The problem that $\lambda_m>1$ occurs in none of these transfers. 
In this step the bang-bang control is not solved. Instead, the homotopic approach is still used, and the perturbation added to the performance index is in the form of a logarithmic barrier \citep{bertrand2002new} with a $\varepsilon$ of 0.01. 
Second, these adjoint variables are scaled to provide an initial guess for the next step where the whole sequence is solved. 
Third, an approximate solution to the bang-bang control is obtained with the logarithmic barrier with a $\varepsilon$ of 0.01. 
Finally, the approximate result is used as the initial guess to solve the bang-bang control directly. 

The overall results of the three sequences are listed in Table~\ref{tab:allseqresults}. 
The subscript $I$ and $R$ denote the initial and refined results, respectively; and $m_f$ denotes the final mass of the probe when the mission is completed. 
The sum of the final mass improves about 5\%, which improves the ranking by one. The improvements in the final masses demonstrate that these methods can be applied to obtain the fuel-optimal multi-asteroid trajectory. 
For all the sequences all the computations are finished in less than 2 seconds, which demonstrates the efficiency of these methods. 

\begin{table}[htbp]
	\caption{Initial and Refined Final Masses of All Three Sequences}
	\label{tab:allseqresults}
	\centering
	\begin{tabular}{c | c | c | c | c} 
		\hline
		 & Seq. 1 & Seq. 2 & Seq. 3 & Sum\\
		\hline
		${m_f}_I$ (kg) & 842.0 & 808.2 & 806.8 & 2457.0\\
		${m_f}_R$ (kg)& 881.3 & 850.9 & 852.1 & 2584.3\\
		Improvement & 4.7\%& 5.3\%& 5.6\% & 5.2\% \\
		\hline
	\end{tabular}
\end{table}

The details of the three sequences are listed in Table~\ref{tab:seq1results}--\ref{tab:seq3results} in APPENDIX B. The classical orbital elements of every asteroid in every sequence are listed in APPENDIX C. 
In Table~\ref{tab:seq1results}--\ref{tab:seq3results} the first column is the name of the asteroid in the sequence. 
The second and third columns list the initial epoch and the corresponding mass when the probe encounters every asteroid, respectively. These results are obtained by the intuitive method. 
The fourth and fifth columns list the refined epoch and the corresponding mass, respectively. 

\begin{figure}[htbp]
	\label{fig:TimeThrustLogBangSC1}
	\centering
	\includegraphics[width=\columnwidth]{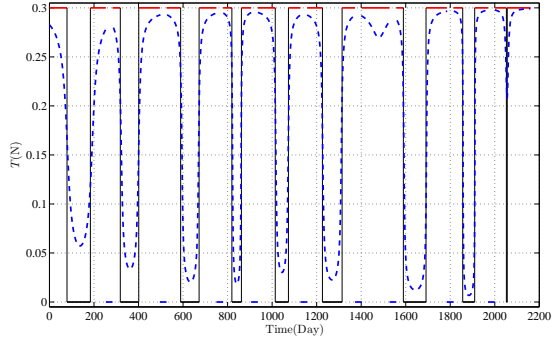}
	\caption{Thrust Magnitude History of Probe 1. Blue: with Logarithmic Homotopy; Red and Black: the Bang-Bang Control}
\end{figure}

As an example, the history of the thrust magnitude of probe 1 is shown in Figure~\ref{fig:TimeThrustLogBangSC1}. 
The blue dashed curves denote the thrust when the homotopic approach is applied. 
The red and black curves denote the thrusting and coasting segments when the bang-bang control is solved, respectively. 
The blue solid lines at the bottom denote when the probe stays at the asteroid. 
When the homotopic approach is used, the thrust magnitude is continuous and is an approximate solution to the bang-bang control. 
The switching detection method demonstrates its ability in dealing with the bang-bang control. 
Two types of thrust structure exist. The first type is composed of two thrusting segments separated by one coasting segment. The second type contains a single thrusting segment, which is easily trimmed in the preliminary design where Lambert problems are solved when approximating low-thrust trajectories by two-impulse trajectories. 

Through checking the value of $\lambda_m$ a posteriori we find that $\lambda_m$ never exceeds unity during the whole mission, which agrees with the previous check that $\lambda_m>1$ occurs in none of the single-leg transfers. 
We might as well propose a conjecture that in similar problems which are composed of multiple single-leg transfers, $\lambda_m$ never exceeds unity from the second leg. 
Although difficult to prove, this condition is necessary if the adjoint scaling technique can be used. 
We might as well list some interesting observations about the physical meaning of $\lambda_m$, although a complete understanding deserves more investigation and it is the future work. 
\begin{enumerate}
	\item The variation of the performance index ${{\updelta}} J$ has the term $\lambda_m(t_0){\updelta} m(t_0)$ \citep{Bryson1975}. A positive $\lambda_m(t_0)$ suggests that the $J$ is improved, i.e., the fuel consumption is reduced, if $m(t_0)$ is reduced. 
	This conclusion corresponds to the fact that larger acceleration is preferable to reduce fuel consumption \citep{tang2016Capture}. 
	\item The improvement of $J$ is larger when $\lambda_m(t_0)$ is larger. 
	\item If the adjoint variable $\lambda_m(t_0)$ exceeds unity, the engine must be on at $t_0$ because equation~\eqref{eq:rho} suggests the switching function is negative no matter what other parameters are. 
\end{enumerate}

\begin{figure}[htbp]
	\centering\includegraphics[width=0.8\columnwidth]{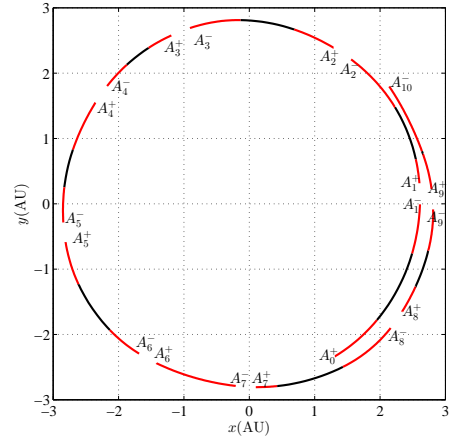}
	\caption{Optimal Trajectory of Probe 1 Projected on the $xy$--Plane}
	\label{fig:BangTrajSC1}
\end{figure}

\begin{figure}[htbp]
	\centering\includegraphics[width=0.8\columnwidth]{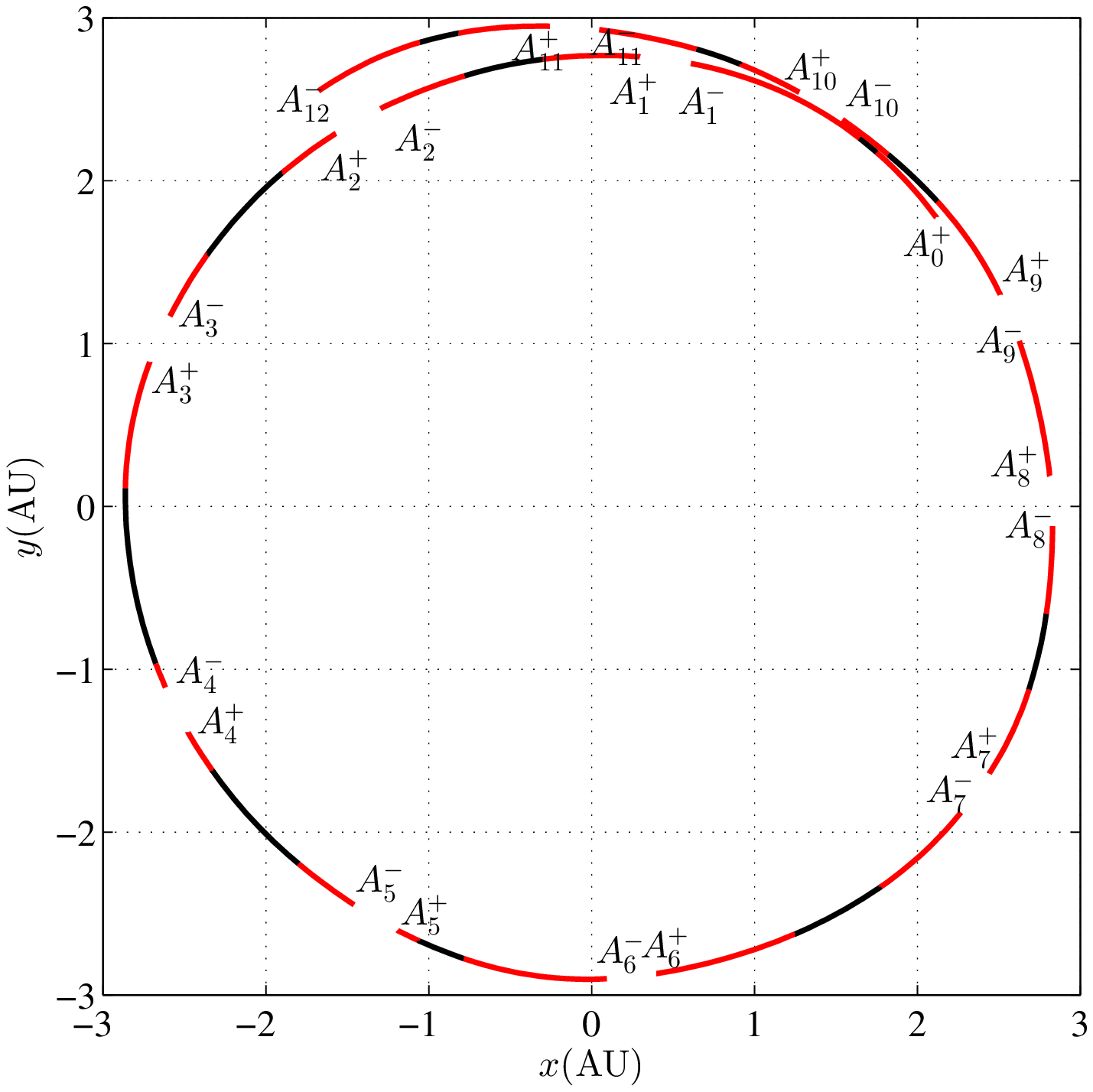}
	\caption{Optimal Trajectory of Probe 2 Projected on the $xy$--Plane}
	\label{fig:BangTrajSC2}
\end{figure}

\begin{figure}[htbp]
	\centering\includegraphics[width=0.8\columnwidth]{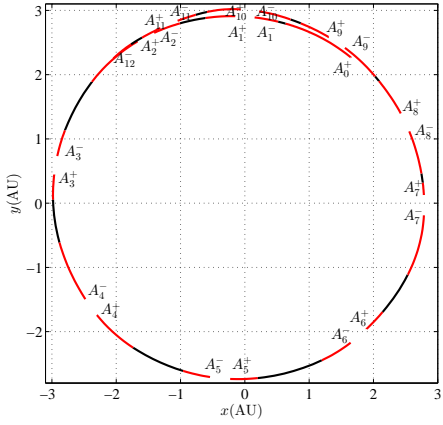}
	\caption{Optimal Trajectory of Probe 3 Projected on the $xy$--Plane}
	\label{fig:BangTrajSC3}
\end{figure}

The optimal trajectories of the three probes are shown in Figures~\ref{fig:BangTrajSC1}--\ref{fig:BangTrajSC3} where red and black arcs denote thrusting and coasting segments, respectively. The discontinuity of the trajectory denotes when spacecraft stays on the asteroid. 

The method of optimizing the whole trajectory has significance. Not only can it be applied to increase the performance index in GTOC, it can also be used to evaluate the largest possible payload as well as the least fuel consumption for deep-space missions. 
Exploration of multiple main-belt asteroids is practical, and thus this method can also be used in the preliminary design to estimate the largest payload. 
When designing the sequence of asteroids, the trim strategy can be less strict in order not to miss the possible global optima. 
As a result the possibility of finding the global optimal solution is higher. 

\section{Conclusion}
The systematic method for optimizing low-thrust trajectories to rendezvous with a dozen main-belt asteroids is proposed. 
The whole trajectories are optimized after introducing dozens of inner-point constraints. 
Indirect methods are rarely applied to optimize such complex trajectories. 
The difficulty arising from the large number of variables is solved, i.e. the difficulty in providing an initial guess is overcome by the adjoint scaling technique, which is the main contribution of this paper.  
The adjoint scaling technique can help provide a tentative guess by solving multiple single-leg transfers, which are usually much easier to solve. 
The adjoint scaling technique connects the adjoint variables of multiple single-leg transfers with those of the multi-leg transfer. 
The fact that this technique may fail if $\lambda_m$ exceeds unity provides a clue to the understanding of its physical meaning. 
Another contribution is that we propose a simple but efficient method for providing initial guesses for transfers between close low-eccentricity and low-inclination orbits. 
The initial guesses of some adjoint variables are obtained by analytically solving a simplified energy-optimal transfer. 
Compared with the random guess, such a guess is more probable and faster to converge and should always be the first choice. 
The improvement in the performance index and the short computation time in numerical examples demonstrate the robustness and efficiency of these methods. 
These methods can be easily extended to other problems which are also composed of multiple legs and every leg is relatively easy to solve. 
However, a better way for handling the cases where $\lambda_m$ exceeds unity and a more general method for providing initial guesses still deserve further investigation. 

\section{Acknowledgment}
This work is supported by the National Natural Science Foundation of China (Grant No. 11302112 and 11432001). The authors thank the reviewers and the editor for their helpful comments and suggestions, especially on the physical meanings of some adjoint variables. 
\appendix
\section*{Appendix: A}
Denote $\boldsymbol{N}=\boldsymbol{M'}\boldsymbol{M'}^{\rm{T}}$ whose components are
\begin{equation}
{\boldsymbol{N}} = \frac{P}{{{\mu}}}\left[ {\begin{array}{*{20}{c}}
	{{{\boldsymbol{N}}_{11}}}&\boldsymbol{0}&\boldsymbol{0}\\
	\boldsymbol{0}&{{{\boldsymbol{N}}_{22}}}&\boldsymbol{0}\\
	\boldsymbol{0}&\boldsymbol{0}&0
	\end{array}} \right]
\end{equation}
where $P=p'$ and
\begin{equation}
{{{\boldsymbol{N}}_{11}}}=\left[ {\begin{array}{*{20}{c}}
	{4{P^2}}&{4P\cos L}&{4P\sin L}\\
	{4P\cos L}&{1 + 3{{\cos }^2}L}&{3\sin L\cos L}\\
	{4P\sin L}&{3\sin L\cos L}&{1 + 3{{\sin }^2}L}
	\end{array}} \right]
\end{equation}
\begin{equation}
{{{\boldsymbol{N}}_{22}}}=\left[ {\begin{array}{*{20}{c}}
	{{{\cos }^2}L}&{\sin L\cos L}\\
	{\sin L\cos L}&{{{\sin }^2}L}
	\end{array}} \right]
\end{equation}

Denote $L_t = L_0 + \omega t$, the indefinite integral of $\boldsymbol{N}_{11}, \boldsymbol{N}_{22}$ with respect to time is
\begin{equation}
{\boldsymbol{N}'_{11}} = \frac{P}{{\omega {\mu}}}\left[ {\begin{array}{*{20}{c}}
	{4\omega {P^2}t}&{4P\sin {L_t}}&{ - 4P\cos {L_t}}\\
	{4P\sin {L_t}}&{\dfrac{3}{2}\sin {L_t}\cos {L_t} + \dfrac{5}{2}{L_t}}&{\dfrac{3}{2}{{\sin }^2}{L_t}}\\
	{ - 4P\cos {L_t}}&{\dfrac{3}{2}{{\sin }^2}{L_t}}&{\dfrac{5}{2}{L_t} - \dfrac{3}{2}\sin {L_t}\cos {L_t}}
	\end{array}} \right]
\end{equation}
\begin{equation}
{{\boldsymbol{N}'}_{22}} = \frac{P}{{8\omega {\mu}}}\left[ {\begin{array}{*{20}{c}}
	{{L_t} + \sin {L_t}\cos {L_t}}&{{{\sin }^2}{L_t}}\\
	{{{\sin }^2}{L_t}}&{{L_t} - \sin {L_t}\cos {L_t}}
	\end{array}} \right]
\end{equation}
\section*{Appendix: B}
\begin{table*}[htbp]
	\caption{Initial and Refined Results of Sequence 1}
	\label{tab:seq1results}
	\centering
	\begin{tabular}{c | c | c | c | c } 
		\hline
		Ast. Name & ${\rm{MJD}}_I$ & $m_I$ (kg)& ${\rm{MJD}}_R$ &$m_R$ (kg)\\
		\hline
		Grantham & 61444.2 & 2000.0 & 61444.2 & 2000.0\\
		1991 ND7 & 61711.7 & 1855.6 & 61727.4 & 1869.3\\
		1998 TN33 & 61964.4 & 1735.4 & 61970.0 & 1754.9\\
		Karlin & 62222.1 & 1608.3 & 62224.1 & 1630.6\\
		1998 WE12 & 62380.4 & 1535.0 & 62379.9 & 1557.2\\
		Tosamakoto & 62603.1 & 1415.6 & 62601.7 & 1440.6\\
		Hermannbondi & 62832.3 & 1288.4 & 62846.9 & 1328.7\\
		Podobed & 63007.4 & 1184.8 & 63001.9 & 1218.6\\
		Woszczyk & 63250.6 & 1057.9 & 63259.0 & 1108.2\\
		Shcheglov & 63434.4 & 963.4 & 63444.2 & 1018.9\\
		Mogamigawa & 63634.1 & 842.0 & 63634.1 & 881.3\\
		\hline
	\end{tabular}
\end{table*}
\begin{table*}[htbp]
	\caption{Initial and Refined Results of Sequence 2}
	\label{tab:seq2results}
	\centering
	\begin{tabular}{c | c | c | c | c } 
		\hline
		Ast. Name & ${\rm{MJD}}_I$ & $m_I$ (kg)& ${\rm{MJD}}_R$ &$m_R$ (kg)\\
		\hline
		Grantham & 61882.2 & 2000.0 & 61882.2 & 2000.0\\
		1259 T-2 & 62112.2 & 1861.8 & 62091.3 & 1854.3\\
		1999 WJ4 & 62306.0 & 1764.3 & 62276.4 & 1758.4\\
		Mimosa & 62482.1 & 1685.2 & 62456.2 & 1685.5\\
		Arpetito & 62728.6 & 1571.1 & 62686.6 & 1603.4\\
		2000 ET45 & 62867.9 & 1497.0 & 62859.8 & 1543.5\\
		1999 CO16 & 63024.4 & 1414.9 & 63019.5 & 1454.4\\
		1999 XL44 & 63272.5 & 1280.1 & 63264.7 & 1317.7\\
		1998 QU47 & 63456.6 & 1176.7 & 63452.4 & 1220.3\\
		Steffl & 63571.0 & 1111.7 & 63563.7 & 1148.7\\
		Silcher & 63744.3 & 1019.7 & 63733.7 & 1060.0\\
		Alprokhorov & 63898.1 & 927.9 & 63894.1 & 970.4\\
		Mogamigawa & 64087.0 & 808.2 & 64087.0 & 850.9\\
		\hline
	\end{tabular}
\end{table*}
\begin{table*}[htbp]	
	\caption{Initial and Refined Results of Sequence 3}
	\label{tab:seq3results}
	\centering
	\begin{tabular}{c | c | c | c | c } 
		\hline
		Ast. Name & ${\rm{MJD}}_I$ & $m_I$ (kg)& ${\rm{MJD}}_R$ &$m_R$ (kg)\\
		\hline
		Grantham & 61947.2 & 2000.0 & 61947.2 & 2000.0\\
		1998 VD13 & 62175.9 & 1890.2 & 62135.6 & 1870.5\\
		Sinyavskaya & 62344.2 & 1812.2 & 62296.9 & 1790.9\\
		Mayakovsky & 62646.4 & 1655.8 & 62554.8 & 1666.8\\
		1999 AP9 & 62847.9 & 1533.6 & 62795.9 & 1543.8\\
		Bohrmann & 63095.3 & 1399.9 & 63019.4 & 1443.8\\
		1999 CA97 & 63276.3 & 1301.7 & 63231.1 & 1366.3\\
		Silcher & 63466.2 & 1192.5 & 63451.8 & 1256.0\\
		1997 DR & 63584.0 & 1133.2 & 63579.0 & 1182.2\\
		Radishchev & 63742.8 & 1038.1 & 63737.1 & 1076.9\\
		H\~afez & 63885.0 & 956.5 & 63880.1 & 991.5\\
		2000 ET165 & 64020.9 & 876.6 & 64015.7 & 915.0\\
		Mogamigawa & 64138.3 & 806.8 & 64138.3 & 852.1\\
		\hline
	\end{tabular}
\end{table*}
\section*{Appendix: C}
\begin{table}[H]
	\caption{Classical Orbital Elements of the Asteroids in the First Sequence at Epoch MJD 56800}
	\label{tab:orbitselements1}
	\centering
	\begin{tabular}{c | c | c | c | c |c |c} 
		\hline
		Ast. Name & $a$(AU) & $e$ & $i$(deg) &$\omega$(deg)&$\varOmega$(deg)&$M$(deg)\\
		\hline
		Grantham 	 & 2.8351662 & 0.0636863 & 1.2216500 & 21.1984300 & 309.0437500 & 87.6164602\\
		1991 ND7 	&2.7057099	&0.0474122	&2.7967900	&160.1996600	&155.1344800	&29.6144276\\
		1998 TN33 	&	2.8928373	&0.0711471	&3.1684200	&298.4774200	&89.8558800	&67.3416368\\
		Karlin 			&2.8786586	&0.0139888	&3.5152300	&356.9468000	&98.8148700	&357.9468249\\
		1998 WE12 	&	2.9025244&	0.0362789&	3.2817800&	90.6428100&	95.3195900&	285.1736252\\
		Tosamakoto		&2.8358216	&0.0429025	&3.1132600	&321.1444700	&119.9078400	&342.5165325	\\
		Hermannbondi	&2.7849273	&0.0801049	&1.7763800	&247.9127800	&96.2210600	&55.9168827	\\
		Podobed 		&2.7878827	&0.0412754	&0.9289900	&12.0766700	&157.3179400	&218.4845623	\\
		Woszczyk 	&	2.9038493&	0.0585692&	1.4313700&	199.8942000&	204.0671700&	74.8749440	\\
		Shcheglov 		&2.8808048	&0.0571353&	1.0089900&	164.6990200&	262.5768500&	37.7106098	\\
		Mogamigawa 	&	2.7518476&	0.1116506&	3.1007000&	308.0214700&	347.1361900&	56.7308881	\\
		\hline
	\end{tabular}
\end{table}
\begin{table}[H]
	\caption{Classical Orbital Elements of the Asteroids in the Second Sequence at Epoch MJD 56800}
	\label{tab:orbitselements2}
	\centering
	\begin{tabular}{c | c | c | c | c |c |c} 
		\hline
		Ast. Name & $a$(AU) & $e$ & $i$(deg) &$\omega$(deg)&$\varOmega$(deg)&$M$(deg)\\
		\hline
		Grantham&2.83517&0.06369&1.22165&21.19843&309.04375&87.61646\\
		1259 T-2&2.89408&0.04969&2.35270&90.45222&31.86648&339.75347\\
		1999 WJ4&2.85203&0.04612&1.53399&70.69924&355.88393&6.65304\\
		Mimosa&2.87398&0.04722&1.17803&108.28712&329.40022&8.57038\\
		Arpetito&2.87457&0.01452&1.02431&305.81295&297.62030&210.19707\\
		2000 ET45&2.93180&0.05444&1.72475&178.14692&359.74187&306.18772\\
		1999 CO16&2.95045&0.02487&2.16902&295.50305&27.34924&181.62682\\
		1999 XL44&2.83978&0.03896&1.12783&159.71372&323.00591&307.57937\\
		1998 QU47&2.85584&0.04394&1.69997&249.47639&189.61051&4.84867\\
		Steffl&2.82693&0.03731&2.16367&100.49009&195.07442&118.81004\\
		Silcher&2.95964&0.05885&0.46464&246.80145&125.27473&138.18915\\
		Alprokhorov&2.99640&0.10778&2.93522&281.08595&84.17199&163.66004\\
		Mogamigawa&2.75185&0.11165&3.10070&308.02147&347.13619&56.73089\\
		\hline
	\end{tabular}
\end{table}
\begin{table}[H]
	\caption{Classical Orbital Elements of the Asteroids in the Third Sequence at Epoch MJD 56800}
	\label{tab:orbitselements3}
	\centering
	\begin{tabular}{c | c | c | c | c |c |c} 
		\hline
		Ast. Name & $a$(AU) & $e$ & $i$(deg) &$\omega$(deg)&$\varOmega$(deg)&$M$(deg)\\
		\hline
		Grantham	&2.83517&0.06369&1.22165&21.19843&309.04375&87.61646\\
		1998 VD13	&2.89350&0.06760&3.24541&307.42881&44.90694&98.51234\\
		Sinyavskaya	&2.87595&0.07705&2.70407&298.61193&50.22626&91.28100\\
		Mayakovsky	&2.87651&0.05619&2.21570&287.00774&25.34998&134.58903\\
		1999 AP9	&2.94598&0.03595&1.57970&82.83970&191.15193&211.96557\\
		Bohrmann	&2.85353&0.05767&1.81556&132.91074&184.53332&115.08813\\
		1999 CA97	&2.96704&0.08180&2.11304&151.83228&162.60897&193.57845\\
		Silcher	&2.95964&0.05885&0.46464&246.80145&125.27473&138.18915\\
		1997 DR	&2.79916&0.02324&2.76067&268.16626&31.24239&94.84897\\
		Radishchev	&2.87780&0.06612&1.33509&347.22497&336.83005&125.02051\\
		H\~afez	&2.84585&0.09951&1.73301&293.91723&25.51876&103.16738\\
		2000 ET165	&2.92012&0.03646&2.05960&276.41478&16.21004&192.18541\\
		Mogamigawa	&2.75185&0.11165&3.10070&308.02147&347.13619&56.73089\\
		\hline
	\end{tabular}
\end{table}
\bibliographystyle{spr-mp-nameyear-cnd}
\bibliography{references}

\begin{thebibliography}{21}
\ifx \bisbn   \undefined \def \bisbn  #1{ISBN #1}\fi
\ifx \binits  \undefined \def \binits#1{#1} \fi
\ifx \bauthor  \undefined \def \bauthor#1{#1} \fi
\ifx \batitle  \undefined \def \batitle#1{#1} \fi
\ifx \bjtitle  \undefined \def \bjtitle#1{#1}\fi
\ifx \bvolume  \undefined \def \bvolume#1{\textbf{#1}}\fi
\ifx \byear  \undefined \def \byear#1{#1} \fi
\ifx \bissue  \undefined \def \bissue#1{#1} \fi
\ifx \bfpage  \undefined \def \bfpage#1{#1} \fi
\ifx \blpage  \undefined \def \blpage #1{#1} \fi
\ifx \burl  \undefined \def \burl#1{\textsf{#1}} \fi
\ifx \doiurl  \undefined \def \doiurl#1{\textsf{#1}} \fi
\ifx \betal  \undefined \def \betal{\textit{et al.}} \fi
\ifx \binstitute  \undefined \def \binstitute#1{#1} \fi
\ifx \binstitutionaled  \undefined \def \binstitutionaled#1{#1} \fi
\ifx \bctitle  \undefined \def \bctitle#1{#1} \fi
\ifx \beditor  \undefined \def \beditor#1{#1} \fi
\ifx \bpublisher  \undefined \def \bpublisher#1{#1} \fi
\ifx \bbtitle  \undefined \def \bbtitle#1{#1} \fi
\ifx \bedition  \undefined \def \bedition#1{#1} \fi
\ifx \bseriesno  \undefined \def \bseriesno#1{#1} \fi
\ifx \blocation  \undefined \def \blocation#1{#1} \fi
\ifx \bsertitle  \undefined \def \bsertitle#1{#1} \fi
\ifx \bsnm \undefined \def \bsnm#1{#1} \fi
\ifx \bsuffix \undefined \def \bsuffix#1{#1} \fi
\ifx \bparticle \undefined \def \bparticle#1{#1} \fi
\ifx \barticle \undefined \def \barticle#1{#1} \fi
\ifx \bconfdate \undefined \def \bconfdate #1{#1} \fi
\ifx \botherref \undefined \def \botherref #1{#1} \fi
\ifx \url \undefined \def \url#1{\textsf{#1}} \fi
\ifx \bchapter \undefined \def \bchapter#1{#1} \fi
\ifx \bbook \undefined \def \bbook#1{#1} \fi
\ifx \bcomment \undefined \def \bcomment#1{#1} \fi
\ifx \oauthor \undefined \def \oauthor#1{#1} \fi
\ifx \citeauthoryear \undefined \def \citeauthoryear#1{#1} \fi
\ifx \endbibitem  \undefined \def \endbibitem {}\fi
\ifx \bconflocation  \undefined \def \bconflocation#1{#1} \fi
\ifx \arxivurl  \undefined \def \arxivurl#1{\textsf{#1}} \fi

\bibitem[\protect\citeauthoryear{Bertrand and Epenoy}{2002}]{bertrand2002new}
\begin{barticle}
\bauthor{\bsnm{Bertrand}, \binits{R.}},
\bauthor{\bsnm{Epenoy}, \binits{R.}}:
\bjtitle{Optimal Control Applications and Methods}
\bvolume{23}(\bissue{4}),
\bfpage{171}
(\byear{2002})
\end{barticle}
\endbibitem

\bibitem[\protect\citeauthoryear{Bryson and Ho}{1975}]{Bryson1975}
\begin{bbook}
\bauthor{\bsnm{Bryson}, \binits{A.E.}},
\bauthor{\bsnm{Ho}, \binits{Y.C.}}:
\bbtitle{Applied Optimal Control; Optimization, Estimation, and Control}.
\bpublisher{Hemisphere},
\blocation{Washington}
(\byear{1975})
\end{bbook}
\endbibitem

\bibitem[\protect\citeauthoryear{Casalino}{2014}]{casalino2014approximate}
\begin{barticle}
\bauthor{\bsnm{Casalino}, \binits{L.}}:
\bjtitle{Journal of Guidance, Control, and Dynamics}
\bvolume{37}(\bissue{3}),
\bfpage{1003}
(\byear{2014})
\end{barticle}
\endbibitem

\bibitem[\protect\citeauthoryear{Casalino and
  Simeoni}{2012}]{casalino2012indirect}
\begin{barticle}
\bauthor{\bsnm{Casalino}, \binits{L.}},
\bauthor{\bsnm{Simeoni}, \binits{F.}}:
\bjtitle{Journal of Guidance, Control, and Dynamics}
\bvolume{35}(\bissue{2}),
\bfpage{423}
(\byear{2012})
\end{barticle}
\endbibitem

\bibitem[\protect\citeauthoryear{Casalino et~al.}{2007}]{casalino20071st}
\begin{barticle}
\bauthor{\bsnm{Casalino}, \binits{L.}},
\bauthor{\bsnm{Colasurdo}, \binits{G.}},
\bauthor{\bsnm{Sentinella}, \binits{M.R.}}:
\bjtitle{Acta Astronautica}
\bvolume{61}(\bissue{9}),
\bfpage{769}
(\byear{2007})
\end{barticle}
\endbibitem

\bibitem[\protect\citeauthoryear{Casalino et~al.}{2014}]{CasalinoGTOC5}
\begin{barticle}
\bauthor{\bsnm{Casalino}, \binits{L.}},
\bauthor{\bsnm{Pastrone}, \binits{D.}},
\bauthor{\bsnm{Simeoni}, \binits{F.}},
\bauthor{\bsnm{Colasurdo}, \binits{G.}},
\bauthor{\bsnm{Zavoli}, \binits{A.}}:
\bjtitle{Acta Futura}
\bvolume{8},
\bfpage{29}
(\byear{2014})
\end{barticle}
\endbibitem

\bibitem[\protect\citeauthoryear{Gao and Kluever}{2004}]{GaoKluever-409}
\begin{botherref}
\oauthor{\bsnm{Gao}, \binits{Y.}},
\oauthor{\bsnm{Kluever}, \binits{C.A.}}:
No. AIAA
\textbf{5088}
(2004)
\end{botherref}
\endbibitem

\bibitem[\protect\citeauthoryear{Gatto and Casalino}{2015}]{gatto2015fast}
\begin{botherref}
\oauthor{\bsnm{Gatto}, \binits{G.}},
\oauthor{\bsnm{Casalino}, \binits{L.}}:
Journal of Guidance, Control, and Dynamics
(2015)
\end{botherref}
\endbibitem

\bibitem[\protect\citeauthoryear{Gong and Li}{2015}]{gong2015equilibria}
\begin{barticle}
\bauthor{\bsnm{Gong}, \binits{S.}},
\bauthor{\bsnm{Li}, \binits{J.}}:
\bjtitle{Astrophysics and Space Science}
\bvolume{355}(\bissue{2}),
\bfpage{213}
(\byear{2015})
\end{barticle}
\endbibitem

\bibitem[\protect\citeauthoryear{Jiang et~al.}{2012}]{jiang2012}
\begin{barticle}
\bauthor{\bsnm{Jiang}, \binits{F.}},
\bauthor{\bsnm{Baoyin}, \binits{H.}},
\bauthor{\bsnm{Li}, \binits{J.}}:
\bjtitle{Journal of Guidance, Control, and Dynamics}
\bvolume{35}(\bissue{1}),
\bfpage{245}
(\byear{2012})
\end{barticle}
\endbibitem

\bibitem[\protect\citeauthoryear{Jiang et~al.}{2014}]{JiangGTOC5}
\begin{barticle}
\bauthor{\bsnm{Jiang}, \binits{F.}},
\bauthor{\bsnm{Chen}, \binits{Y.}},
\bauthor{\bsnm{Liu}, \binits{Y.}},
\bauthor{\bsnm{Baoyin}, \binits{H.}},
\bauthor{\bsnm{Li}, \binits{J.}}:
\bjtitle{Acta Futura}
\bvolume{8},
\bfpage{37}
(\byear{2014})
\end{barticle}
\endbibitem

\bibitem[\protect\citeauthoryear{Li and Xi}{2012}]{li2012fuel}
\begin{barticle}
\bauthor{\bsnm{Li}, \binits{J.}},
\bauthor{\bsnm{Xi}, \binits{X.-n.}}:
\bjtitle{Journal of Guidance, Control, and Dynamics}
\bvolume{35}(\bissue{6}),
\bfpage{1709}
(\byear{2012})
\end{barticle}
\endbibitem

\bibitem[\protect\citeauthoryear{McInnes}{2002}]{mcinnes2002astronomical}
\begin{barticle}
\bauthor{\bsnm{McInnes}, \binits{C.R.}}:
\bjtitle{Astrophysics and space science}
\bvolume{282}(\bissue{4}),
\bfpage{765}
(\byear{2002})
\end{barticle}
\endbibitem

\bibitem[\protect\citeauthoryear{McKay et~al.}{2011}]{mckay2011survey}
\begin{barticle}
\bauthor{\bsnm{McKay}, \binits{R.}},
\bauthor{\bsnm{Macdonald}, \binits{M.}},
\bauthor{\bsnm{Biggs}, \binits{J.}},
\bauthor{\bsnm{McInnes}, \binits{C.}}:
\bjtitle{Journal of Guidance, Control, and Dynamics}
\bvolume{34}(\bissue{3}),
\bfpage{645}
(\byear{2011})
\end{barticle}
\endbibitem

\bibitem[\protect\citeauthoryear{Rayman et~al.}{2007}]{DAWN}
\begin{barticle}
\bauthor{\bsnm{Rayman}, \binits{M.D.}},
\bauthor{\bsnm{Fraschetti}, \binits{T.C.}},
\bauthor{\bsnm{Raymond}, \binits{C.A.}},
\bauthor{\bsnm{Russell}, \binits{C.T.}}:
\bjtitle{Acta Astronautica}
\bvolume{60}(\bissue{10}),
\bfpage{930}
(\byear{2007})
\end{barticle}
\endbibitem

\bibitem[\protect\citeauthoryear{Tang and Jiang}{2016}]{tang2016Capture}
\begin{barticle}
\bauthor{\bsnm{Tang}, \binits{G.}},
\bauthor{\bsnm{Jiang}, \binits{F.}}:
\bjtitle{Astrophysics and Space Science}
\bvolume{361}(\bissue{1}),
\bfpage{1}
(\byear{2016})
\end{barticle}
\endbibitem

\bibitem[\protect\citeauthoryear{Walker et~al.}{1985}]{walker1985set}
\begin{barticle}
\bauthor{\bsnm{Walker}, \binits{M.}},
\bauthor{\bsnm{Ireland}, \binits{B.}},
\bauthor{\bsnm{Owens}, \binits{J.}}:
\bjtitle{Celestial Mechanics}
\bvolume{36}(\bissue{4}),
\bfpage{409}
(\byear{1985})
\end{barticle}
\endbibitem

\bibitem[\protect\citeauthoryear{Wu et~al.}{2014}]{wu2014artificial}
\begin{barticle}
\bauthor{\bsnm{Wu}, \binits{Z.}},
\bauthor{\bsnm{Jiang}, \binits{F.}},
\bauthor{\bsnm{Li}, \binits{J.}}:
\bjtitle{Astrophysics and Space Science}
\bvolume{352}(\bissue{2}),
\bfpage{503}
(\byear{2014})
\end{barticle}
\endbibitem

\bibitem[\protect\citeauthoryear{Yang et~al.}{2015}]{Yang2015837}
\begin{barticle}
\bauthor{\bsnm{Yang}, \binits{H.}},
\bauthor{\bsnm{Li}, \binits{J.}},
\bauthor{\bsnm{Baoyin}, \binits{H.}}:
\bjtitle{Advances in Space Research}
\bvolume{56}(\bissue{5}),
\bfpage{837}
(\byear{2015})
\end{barticle}
\endbibitem

\bibitem[\protect\citeauthoryear{Zeng et~al.}{2011}]{zeng2011new}
\begin{barticle}
\bauthor{\bsnm{Zeng}, \binits{X.-Y.}},
\bauthor{\bsnm{Baoyin}, \binits{H.}},
\bauthor{\bsnm{Li}, \binits{J.-F.}},
\bauthor{\bsnm{Gong}, \binits{S.-P.}}:
\bjtitle{Research in Astronomy and Astrophysics}
\bvolume{11}(\bissue{7}),
\bfpage{863}
(\byear{2011})
\end{barticle}
\endbibitem

\bibitem[\protect\citeauthoryear{Zeng et~al.}{2014}]{zeng2014fast}
\begin{barticle}
\bauthor{\bsnm{Zeng}, \binits{X.}},
\bauthor{\bsnm{Gong}, \binits{S.}},
\bauthor{\bsnm{Li}, \binits{J.}}:
\bjtitle{Acta Astronautica}
\bvolume{105}(\bissue{1}),
\bfpage{40}
(\byear{2014})
\end{barticle}
\endbibitem

\end{thebibliography}

\end{document}